\documentclass[%
 aip,
amsmath,amssymb,
 reprint,%
]{revtex4-1}

\usepackage{amsthm, amsmath}
\usepackage{eucal}
\usepackage{amssymb}
\usepackage{mathrsfs}
\usepackage{stackengine}
\usepackage{scalerel}
\usepackage{fp}
\usepackage{diagbox}
\usepackage{natbib}
\usepackage{siunitx}
\usepackage{graphicx}
\usepackage{dcolumn}
\usepackage{bm}
\usepackage{diagbox,subfigure}
\usepackage{subcaption,graphicx,ulem}

\usepackage[utf8]{inputenc}
\usepackage[T1]{fontenc}
\usepackage{mathptmx}
\usepackage{etoolbox}
\usepackage{tabularx}

\makeatletter
\def\@email#1#2{%
 \endgroup
 \patchcmd{\titleblock@produce}
  {\frontmatter@RRAPformat}
  {\frontmatter@RRAPformat{\produce@RRAP{*#1\href{mailto:#2}{#2}}}\frontmatter@RRAPformat}
  {}{}
}%

\newcolumntype{Y}{>{\centering\arraybackslash}X}

\makeatother
\begin{document}


\title[Influence of a Xenon interlayer]{Influence of a Xenon interlayer on dissociative electron attachment to deuterated methane  on a platinum substrate}
\author{Norhan OMAR}
\email{norhan.omar@u-bordeaux.fr}
 \affiliation{
 Laboratoire Chrono-Environnement, UMR CNRS 6249, Université de Franche-Comté, 25030 Besançon cedex, France.
 }
 \author{Pierre CLOUTIER}
 \affiliation{
 Department of Nuclear Medicine and Radiobiology, Université de Sherbrooke, Sherbrooke, Quebec J1H 5N4, Canada.
 }
 \author{Christophe RAMSEYER}
 \affiliation{
 Laboratoire Chrono-Environnement, UMR CNRS 6249, Université de Franche-Comté, 25030 Besançon cedex, France.
 }
 \author{Léon SANCHE}
 \affiliation{
 Department of Nuclear Medicine and Radiobiology, Université de Sherbrooke, Sherbrooke, Quebec J1H 5N4, Canada.
 }
 \author{Michel FROMM}
 \affiliation{
 Laboratoire Chrono-Environnement, UMR CNRS 6249, Université de Franche-Comté, 25030 Besançon cedex, France.
 }

\date{\today}

\begin{abstract}
We investigate the impact of intercalating a xenon layer between a thin condensed CD$_4$ film of two monolayers (ML) and a platinum surface on the dissociative electron attachment (DEA). The observed desorption results are compared with density functional theory (DFT) calculations, which reveal the binding energies of various anionic and neutral species as a function of the xenon film thickness on the Pt (111) substrate. The theoretical results suggest that 6 ML of xenon are sufficient to diminish the surface effect, enabling physisorbed anionic fragments to desorb from the CD$_4$ film. In contrast, 20 ML (approximately 10 nm) are experimentally necessary to achieve saturation in the desorption of D$^-$. In addition, the presence of xenon layers enables the coupling of resonance states with Xe excited states, thereby inhibiting the electrons from returning to the metal.
Aside from reducing surface interactions, the xenon interlayer significantly enhances DEA to CD$_4$.

\end{abstract}

\maketitle
Dissociative electron attachment (DEA) occurs via the capture of a low energy electron (LEE), of typically 0-16 eV, into a previously unoccupied orbital of a molecule resulting a dissociative state configuration. A priory, such an unstable transient anion (TA; e.g., AB$^{-}$) can relax by autodetaching an electron (AB$^{-}$ $\rightarrow$ e$^{-}$ + AB), but if its lifetime is sufficiently long, it can dissociate into produce A$^{-}$ or B$^{-}$ and the corresponding radical. First revealed in the 1960s \cite{Craggs1960,HICKAM1959458}, DEA has been since observed for most molecules investigated, although its cross section varies over orders of magnitude. In that respect, mass spectrometry has contributed considerably to the detection of the anionic products resulting in the decay of TAs into the DEA channel. In those experiments electron stimulated desorption (ESD) yields are measured as a function of incident electron energy to obtain a yield function. Classical diatomic molecules like H$_{2}$ \cite{schulz1959,Rapp1965,Krishnakumar2011} and O$_{2}$ \cite{Prabhudesai2006} were the first to be investigated for electron attachment, leading to the production of H$^{-}$ and O$^{-}$ respectively. Subsequently, studies were conducted on alkali halides HX and halogens X$_{2}$ (X = Cl, Br, F), as well as CO \cite{Chantry1968, Stamatovic1970,abouaf1981}. Unsurprisingly, in the case of CO, two different dissociation channels were observed for the production of both O$^{-}$ and, to a lesser extent, C$^{-}$ anions. Furthermore, DEA to CH$_{4}$ has also been studied extensively over the years \cite{Plessis1983,chatham1984, stano2003}. Methane has a very low anion production rate compared to that of positive ions resulting from direct ionization, which occurs at electron energies over 12.6 eV. Stano et al. \cite{stano2003} showed that CH$_4^+$ and CH$_3^+$ cations are mainly detected at 293 K while small amounts of CH$_2^+$, CH$^+$ and CH$_4^+$ and C$^+$ can be also observed together with these two cations at 693 K. This study of Stano and coworkers assumed that only H$^-$ anions was produced. DEA of small halogenated molecules such as CF$_4$ and SF$_6$ have been intensively studied because of their role in chemistry. Interestingly, DEA of CF$_4$ \cite{Fogle20152,omarsson2014,Xia2013} promotes the formation of F$^-$ with a strong DEA cross-section much larger than that of methane. DEA is not restricted to small molecules and appears to be universal. It occurs in much larger molecules than those mentioned above, such as DNA fragments \cite{Pan2006,Aflatooni2006,Konig2006,Ptasiifmmode2007, Kumar2007, Ptasinska2007,Muftakhov2021} and even cellular DNA \cite{Pan2003}. In 2000, Sanche and coworkers \cite{Boudaiffa2000} showed experimentally that TA formation can damage plasmid DNA by causing single-strand and double-strand breaks. It was later established that DEA was involved in the mechanism inducing the damage \cite{Pan2003}.  \\

The DEA cross section strongly depends on the environment of the target molecules. Fabrikant et al. \cite{fabrikant2016} have recently reviewed the effect of surfaces and clusters on DEA to condensed molecules. Through many examples, they showed that the surface environment can lead both to enhancement or suppression of DEA cross sections in addition to shifts in appearance and resonance energies. According to their point of view, enhanced cross sections are typically due to polarization interactions between the TA and the surface, which lead to both an increase in the electron capture cross section and in the survival probability of the TA. For instance, CF$_3$I molecules show an increase by more than two orders of magnitude in the production of CF$_3^-$ compared with the corresponding gas-phase DEA cross section \cite{leCoat1998}. Decreases in the DEA cross section can arise if a vibrational Feshbach resonance (VFR) \cite{gauyacq1982} is the main contributor to the initial electron capture, which is followed by electron transfer to a valence dissociative state. In this case, the environment suppresses the VFR and the DEA cross section is reduced. As an example, a significant suppression of the VFR, which strongly dominates electron capture by gaseous CH$_3$I, occurs in contrast to other molecules like CH$_3$Cl and CF$_3$Cl, under similar conditions \cite{Fabrikant2011}. In a recent investigation, we studied DEA in thin films of CD$_4$ adsorbed on various surfaces. DEA yield functions not only strongly depended on the type of metallic surfaces (Platinum, Graphite, Gold, Tantalum) onto which CD$_4$ molecules are deposited but also on CD$_4$ coverage. The appearance and resonance energies and the relative cross sections in the yield functions of CD$_4$ varied significantly, in a non-trivial manner, depending on the choice of substrate. D$^-$ was the main anionic fragment detected with some smaller quantities of CD$_n^-$ ions (n=1-3) \cite{OMAR2023}, especially at CD$_4$ coverages higher than 2-3 monolayers (ML). Many reasons can be invoked to explain such differences. First, incident electrons can undergo collisions with the adsorbate and the substrate before reaching the target molecule and attaching to one of its unfilled orbital. These electrons can partially transfer their energy to the dense molecular environment. These collisions as well as auto-attachment can depend on the substrate nature where the molecules of interest are adsorbed. Secondly, any TA is dependent of various interactions with the substrate, which ususally decrease its lifetime. The adsorption process not only affects the target molecule and its TAs, but also, the dissociating stable anionic fragments. In the case of CD$_4$ molecules, we have recently showed via ab initio calculations that for CD$_4$ molecules undergoing DEA, D$^-$ and CD$_3^-$ fragments are physisorbed on a platinum (111) surface, while CD$_4^-$ and CD$_n^-$ ions (n=1-2) are chemisorbed at very low coverage (e.g., 1 ML). For these latter species, there is no doubt that electronic and vibrational states are modified with respect to those of the gas phase. This large difference in adsorption energy and electron transfer between anionic fragments and Pt could explain the selective desorption spectra assuming that chemisorbed species remain trapped on the surface. In the presence of a metallic surface, the adsorption phenomena play a significant role close to the surface. It's important to note that the ESD yields from ML or sub-ML coverage of a metal substrate do not reflect the DEA cross section occuring on the surface, as the anions produced by DEA may not desorbed in vacuum due to their strong adsorption to the metal. This can occur when the anions stick to the surface with a sufficiently large adsorption energy. To control and reduce adsorption effects, a thin, insulating layer (e.g. a rare gas solid film) can be inserted between the metal substrate and the target molecule. For example, an argon spacer film has been condensed on a Pt substrate to control the vibrational excitation of N$_{2}$ molecules by low energy electron impact, which is dominated by the N$_{2}^{-}$ ($^2\Pi_g$) resonance \cite{Marinica2001}. If the spacer can reduce the adsorption interactions due to the metallic surface, it can also modify the characteristics of the resonances affected by the substrate. 

In the  present work, using the same spacer approach, we condense CD$_4$ molecules on Xe films pre-deposited on a platinum substrate, whose surface presents azimuthally disordered (111) facets \cite{Perluzzo1985}. With such configurations, we investigate the role of molecular adsorption via DEA to CD$_4$. Several thicknesses of xenon (from 1 to 20 ML) are inserted between the Pt surface and CD$_4$ films. ESD experiments are conducted with a specific focus on ESD yield functions of D$^{-}$. Anionic yields and the appearance and resonance energies are recorded as a function of the Xe-spacer thickness. Adsorption properties are also calculated. The comparison between theory and experiments are made to better understand DEA vs adsorption/desorption competition. Experiments are described in Section 2, together with the calculated background yields. The ESD yield functions of D$^{-}$ arising from DEA to CD$_4$ deposited on Pt(111) for several thickness of Xe are presented in section 3. The role of the Xe-spacer is clearly identified. We find that adsorption energies alone cannot explain the role of the spacer on the anionic yields at large Xe thicknesses.

\section{Materials and methods}
\subsection{Experiments}

The apparatus used to measure anion ESD has been described elsewhere \cite{Hedhili2006,Bazin2009}. For this work, nanoscale films of CD$_4$ and Xe were formed by condensation onto a polycrystalline platinum substrate held at a temperature of 18 K in an ultra-high vacuum chamber (base pressure 3 x 10$^{-10}$ torr). After repeated heating around 1500 K, the substrate crystalizes producing microfacets with preferential (111) orientation normal to the surface \cite{Perluzzo1985}. Typically, CD$_4$ films of 2 monolayer (ML) thickness were deposited on Xe films of differing thickness. Such thicknesses were estimated to an absolute accuracy of < 30 $\%$ from the quantity of each sample vapor required to form a 1 ML thick film, as determined by LEE transmission measurements \cite{Bader1982}. A pulsed beam  of LEE (pulse duration = 750 ns at 5kHz)  from a Kimball Physics ELG-2 gun is incident  on the molecular films. The cross-sectional area of the electron beam is approximately 2.5 mm$^2$ and it is incident at an angle of 45 \textdegree relative to the sample normal. The incident current on the sample, averaged over time, is 5 nA.

Desorbed negative ions are pushed into the input optics of a time-of-flight mass spectrometer (Kore-5000 Reflectron TOF analyzer) by applying a -2400 V voltage pulse to the metal substrate after each electron pulse. The variation of each desorbed anion yield with incident electron energy (termed anion yield functions) are obtained from multiple TOF mass spectra (each acquired over 10s) at sequentially increasing incident energies (increments of 0.25 eV), When such measurements are repeated on a single sample and the yield functions are essentially unchanged, it can be concluded that the accumulation of charge during measurement and/or sample degradation, are negligible.

\subsection{Computational methodology} 
\textbf{Calculation of adsorption properties} \\

\begin{figure*}[!ht]
	\centering
	\includegraphics[scale=0.33]{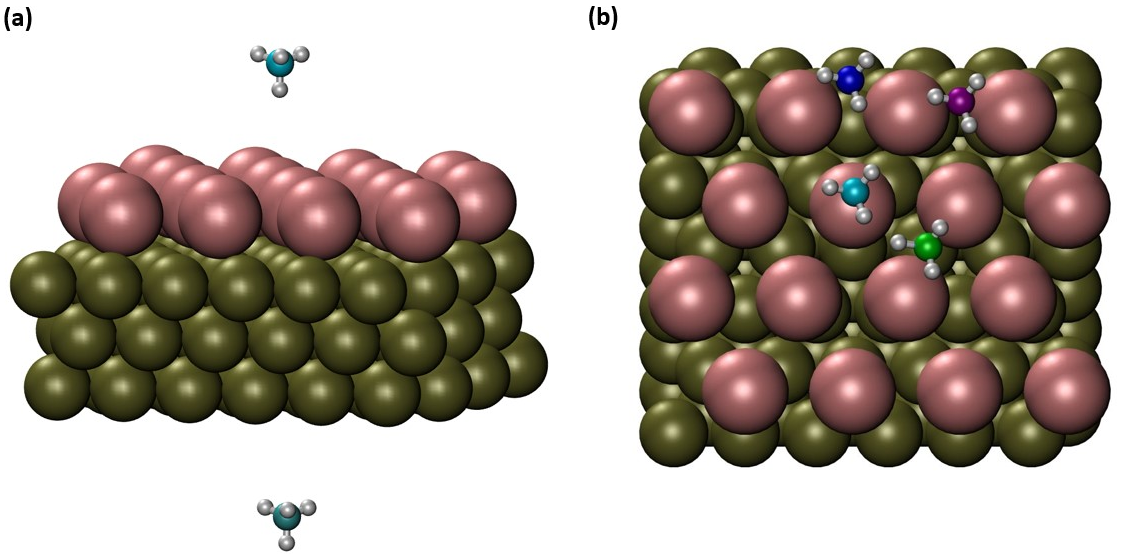}
	\caption{(a) Xenon monolayer coadsorbed on three Pt (111) MLs with two symmetric fragments (CD$_4$) in side view. (b) The four positions of the studied adsorbate relative to the platinum surface in top view: Bridge (cyan), On-top (green), HCP, hexagonal close-packed (purple) and FCC, face centered cubic (dark blue). }
	\label{fig:StructVASP}
\end{figure*}

The adsorption of CD$_4$ on the platinum surface was simulated  from the model depicted in Fig.\ref{fig:StructVASP}. We considered a slab of 3 Pt(111) layers each containing a (6$\times$6) surface unit cell of lattice parameter a$_{Pt}$. Xe(111) layers were placed on top of the Pt slab to mimic the spacer. Four thicknesses of Xe(111) layers (one, two, three, and six layers) were thus coadsorbed on the 3 layers of Pt(111). Each layer of Xe contains (4$\times$4) surface unit cells of lattice parameter $a_{Xe}$. In addition, a 70 Å thick vacuum space was added to the Xe/Pt slab inteface to prohibit artificial interactions between two repeated systems in the normal direction to the surface inherent to the use of periodic boxes. In studying the adsorption of the anion on the surface, a counter-ion was added in the vacuum space at 35 Å from the Xe/Pt slab. This cation (He$^{+}$) which is inversely charged maintains the neutrality of the system. Any undesired electrostatic interaction between the anion and the cation is avoided due to the separation between the two ions. CD$_4$, CD$_4^-$, D and D$^-$ were placed at the interface between the xenon surface and vacuum. 

Density functional theory (DFT) calculations were conducted to optimize the structure of each investigated system. Structure relaxation of all ions and neutral molecules adsorbed on the surface were performed using the Vienna ab initio Simulation Package (VASP) code  \citep{Hafner1997, Kresse1993, Kresse1994, Kresse19966, Kresse199654}. Ionic cores were represented by projector-augmented wave (PAW) pseudopotentials in all calculations \citep{Blochl1994, Kresse1999}. The plane wave cutoff energy used in these calculations was chosen to be 600 eV. The exchange-correlation function of the revised Perdew-Burke-Ernzerhof function (revPBE) \citep{Zhang1998, Perdew1996} was used in combination with the GGA approximation \citep{Autschbach2009, Maitra2017}. The convergence criterion applied on the ionic relaxation was 10$^{-4}$ eV and that on the electronic relaxation was 10$^{-5}$ eV. The Brillouin zone was explored with a (5$\times$5$\times$1) k-points grid with a Fermi level shift of 0.2 eV using the Methfessel-Paxton scheme. The electronic charge optimization was performed using the Davidson iteration method. For physisorbed species, dispersion-repulsion contributions are not well taken into account in DFT calculations. 
Such correction due to long range interactions were introduced via the Grimme DFT+D2 approach. Additionally, a dipole correction was applied by allowing the system to adsorb fragments on the two sides of surface, which avoids the artificial electrostatic interaction induced between the surface dipole moments of the asymmetrically repeated slabs \citep{Thirumalai2016}.
The optimization of the 3 layers of the Pt(111) lead to a$_{Pt}$=4.02 Å. The equilibrium Xe lattice constant was found equal to a$_{Xe}$=6.13 Å for Xe(111) layers. When subsequent molecular adsorption was considered, DFT optimization was conducted only on the molecules and the Xe atoms. The Pt atoms were not allowed to relax during these energy calculations considering that molecular adsorption does not influence substrate species. By contrast, the xenon atoms were not considered fixed in their positions in response to adsorbing molecules or ions. \\  

To derive the adsorption energy of a given X fragment (neutral or anionic), we performed
calculations of the ground state energies of coadsorbed xenon on the Pt(111) slab (E$_{Xe/Pt}$), the single fragment X (E$_{X}$) and the adsorbed X/Xe/Pt(111) coupled systems (E$_{2X/Xe/Pt}$) by applying the following equation:

\begin{equation}
    E_{ads \ (X)}=
    \frac{1}{2} \left ( E_{2X/Xe/Pt} - E_{Xe/Pt}  -  2E_{X} \right) \left \{ 
\begin{array}{l}
Neutral: X \\
Anion: X^- + He^+
\end{array}
\right. 
\label{Eq:Eads}
\end{equation}
These calculations apply to the high symmetry adsorption sites of fragment X, notably Bridge, On-top, HCP and FCC configurations (Fig.\ref{fig:StructVASP}b). \\

\textbf{Calculation of excited states in the gas phase} \\
In this work, DEA to CD$_4$ adsorbed on Pt produces D$^{-}$ and CD$_n^{-}$ (n=1-3) fragments with incident electrons of about 10 eV. Since the spacer is also subjected to the electrons beam, we have also investigated the electronic properties of Xe. We have first performed calculations of the excited states of the atom Xe$^{*}$ in the gas phase. All of these calculations have been performed using the Gaussian 09 software \citep{g09}. First, the xenon atom was optimized in its ground state at the DFT-BLYP/Def2QZVPP level of theory \citep{Ricca1995, Becke1993}. At this optimized geometry, the energy levels of the 40 excited states of Xe$^{*}$ were determined using time-dependent density functional theory (TDDFT) \citep{Runge1984, Petersilka1996, Casida2009} at the same level of theory, BLYP/Def2QZVPP. The identification of the excitation energy levels of Xe is based on the energy difference between the ground state energy of the Xe atom and the excited state energy of Xe$^{*}$.

\section{Results and discussion}
\begin{figure}[h!]
     \centering
        \includegraphics[scale=0.3]{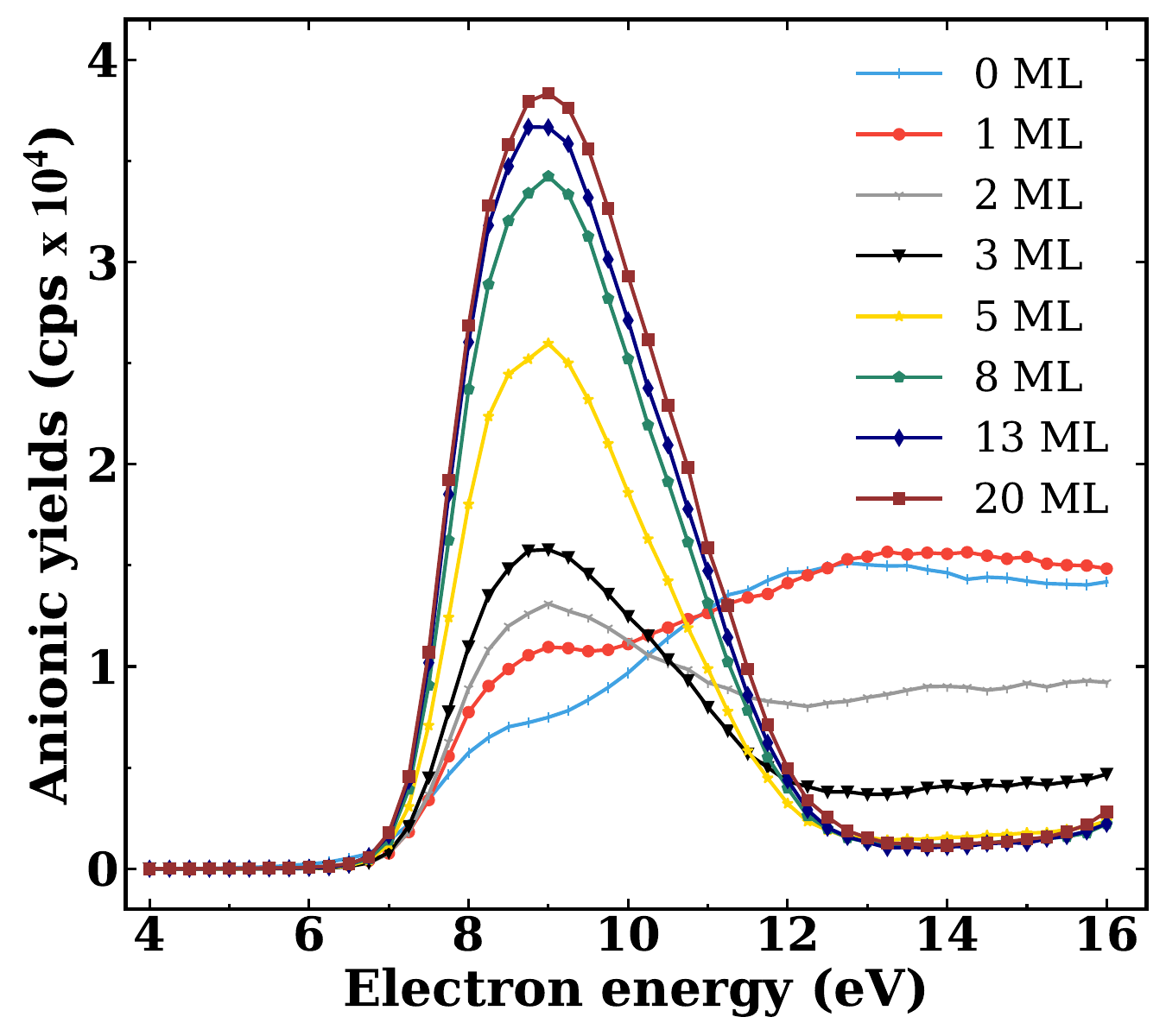}
    \caption{Anion yield function of D$^-$ desorbed from 2 ML of CD$_4$ deposited on top of different layer thicknesses of Xe adsorbed on Pt(111). }
    \label{Fig:Yields_2MLCD4}
\end{figure}

The anions produced by DEA to CD$_4$ detected in the present experiments were D$^-$, CD$^{-}$, CD$_{2}^{-}$ and CD$_{3}^{-}$, with a strong preponderance for D$^-$. In our previous study, without the Xe spacer, D$^{-}$ was also observed to predominantly desorb relative to the other stable anions. Here, the yields of D$^{-}$ desorbed from 2 ML CD$_4$ deposited on the Pt substrate were recorded for Xe spacer thicknesses of 0-3, 5, 8, 13 and 20 ML as shown in Fig.\ref{Fig:Yields_2MLCD4}. A coverage of 2 ML was chosen to be sufficiently thin to minimize the effects of fragment collisions in the film during anion desorption, but thick enough to produce sufficient and reproducible desorption yields of D$^-$. The main characteristics of the resonance peaks (appearance and resonance energies, maximum yield) in the ESD yield functions are given in Table \ref{Table:Yield_2ML}. 

\begin{figure}[!ht]
\centering
\includegraphics[scale=0.33]{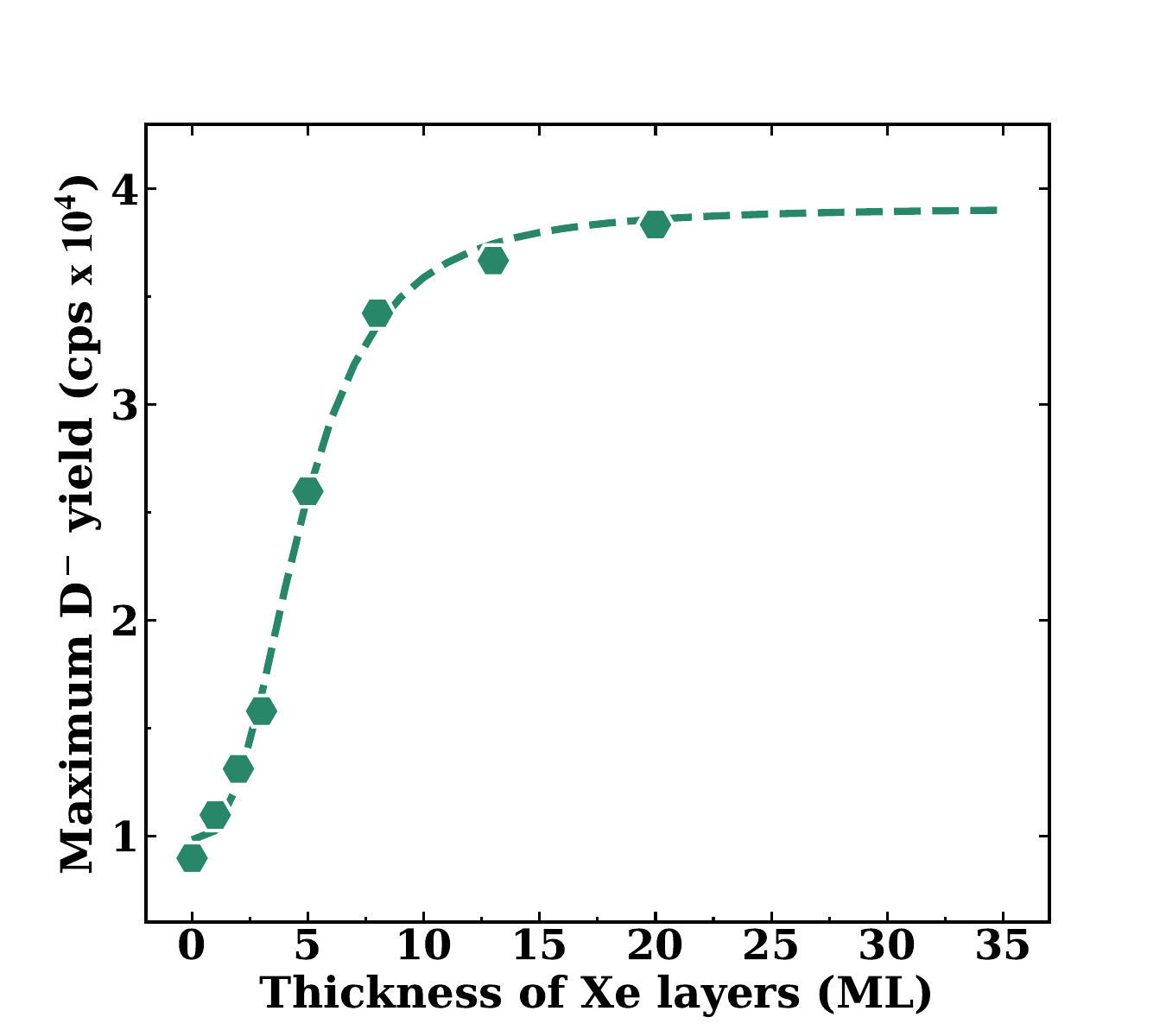}
\caption{Xe film thickness dependence of the maximum yields of D$^-$ desorbed from 2 ML of CD$_4$ deposited on top of Xe multilayers adsorbed on Pt(111). The green hexagons represent the experimental values and the green dashed line represents the fitted curve.}
\label{fig:DYield_Max}
\end{figure}

It is interesting to note that the Xe thickness only affects the maximum yield value, while the spacer has little influence on the appearance and resonance energies, which remain constant at 6.5 and 9 eV, respectively. When the CD$_4$ film is moved away from the platinum surface by 20 ML of Xe, the maximum yield of D$^-$ ions desorbed from the film increases fourfold compared to that desorbed from the CD$_4$ film directly deposited on the bare Pt surface \cite{Omar2024}. The maximum yield of D$^-$ ions grows from 8962 cps at 0 ML of Xe to 38347 cps at 20 ML of Xe. Therefore, the observed intensity differences of the maximum in D$^-$ yield functions can only be explained by the adsorption properties or the role of the spacer. Figure \ref{fig:DYield_Max} represents the maximum yields of D$^-$ desorbed from 2 ML of CD$_4$ as a function of varying xenon thickness, showing that the maximum yield behavior does not depend linearly on the spacer thickness and that saturation occurs for Xe thicknesses greater than about 20 ML.

\begin{table}[ht!]
\caption{\label{Table:Yield_2ML}Resonance energy (RE) and appearance energy (AE) in eV, maximum yield in cps and DEA peak area in 10$^{5}$ cps.eV for D$^-$ desorbed by electron impact on 2 ML of CD$_4$ condensed on Xe films of different thicknesses deposited on a Pt substrate.}
\begin{center}
\begin{tabularx}{0.95\linewidth}{Y|Y|Y|Y|Y}
\hline \hline
\centering
Thickness &  AE & RE   & Max yields & Area \\ 
\hline \hline
0 & 6.5  & 9.0 & 8962 & 1.047  \\ 
1 & 6.5  & 9.0 & 10963 & 1.128  \\ 
2 & 6.5  & 9.0 & 13106  & 0.834  \\  
3 & 6.5  & 9.0 & 15783 & 0.671 \\ 
5 & 6.5  & 9.0 & 25977 & 0.825  \\ 
8 & 6.5  & 9.0 & 34239 & 1.060  \\ 
13 & 6.5 & 9.0 & 36683 & 1.149  \\ 
20 & 6.5 & 9.0 & 38347 & 1.229 \\ 
\hline \hline
\end{tabularx}
\end{center}
\end{table}

\begin{figure*}[!ht]
\centering
\begin{subfigure}
        \centering
        \includegraphics[scale=0.28]{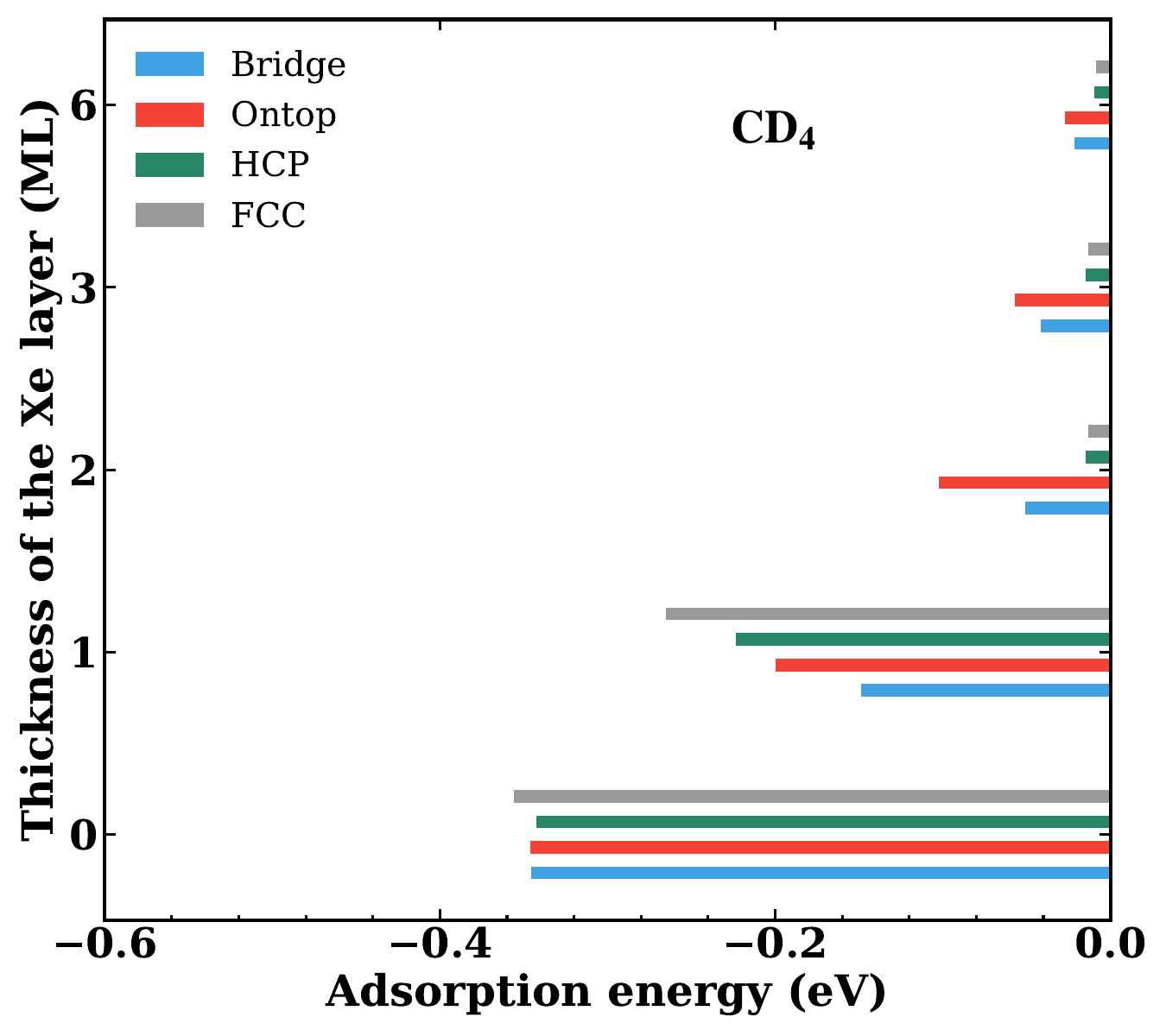}
        \label{Fig:Eads_CD4}
    \end{subfigure}
    \hfill%
    \begin{subfigure}
      \centering
        \includegraphics[scale=0.28]{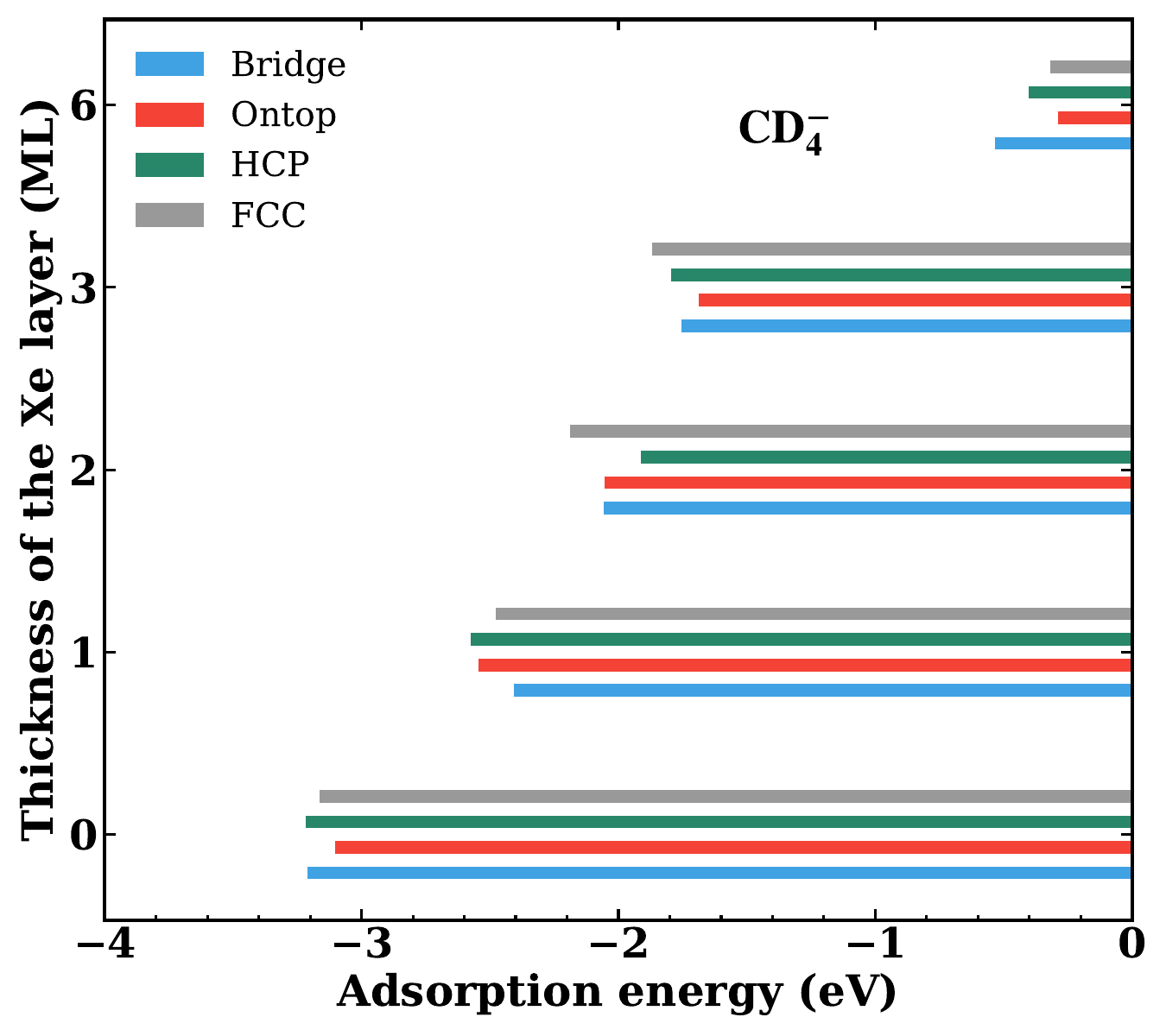}
        \label{Fig:Eads_CD4An}
    \end{subfigure}
    \begin{subfigure}
        \centering
        \includegraphics[scale=0.273]{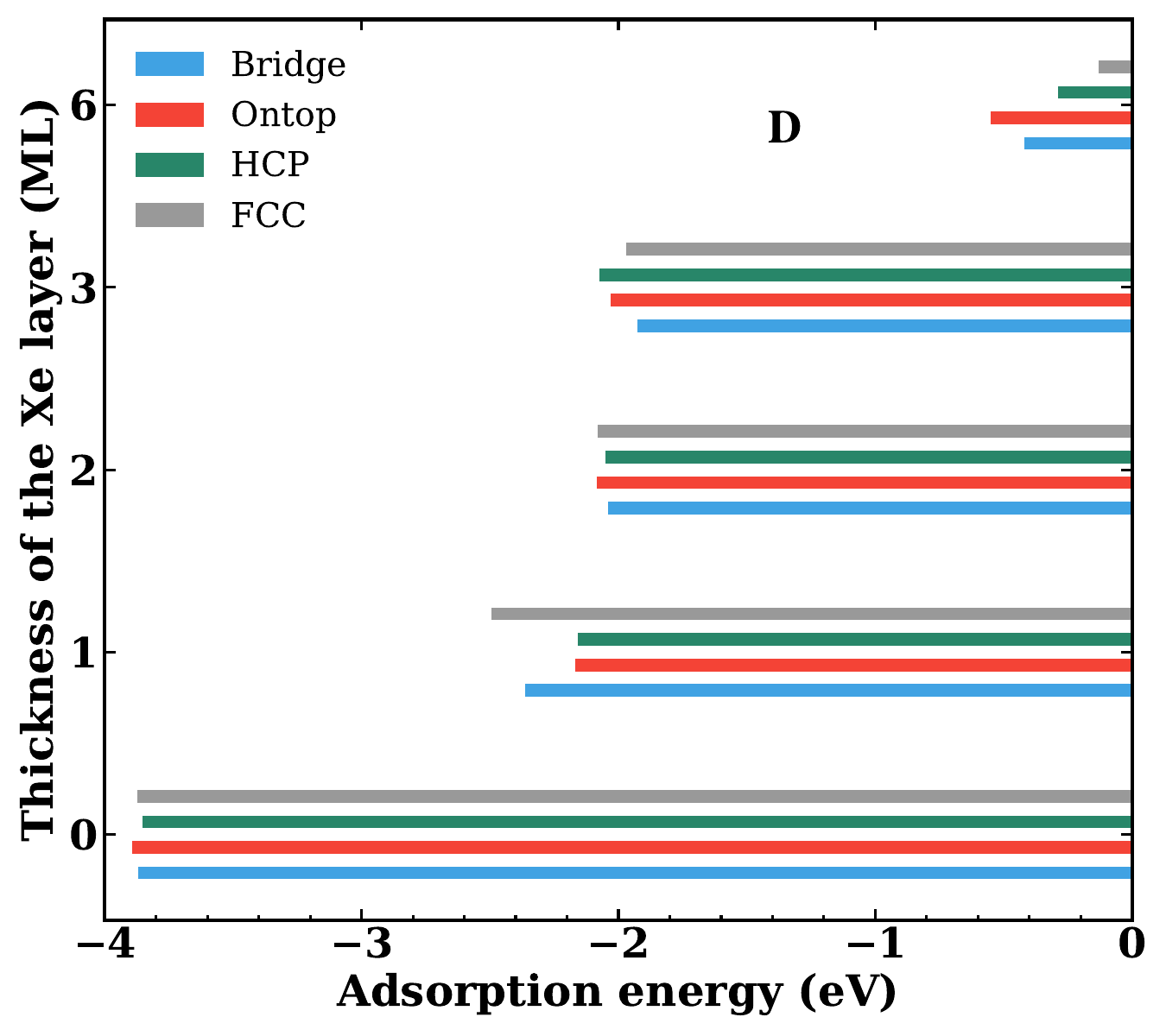}
        \label{Fig:Eads_D}
    \end{subfigure}
    \hfill%
    \begin{subfigure}
        \centering
        \includegraphics[scale=0.275]{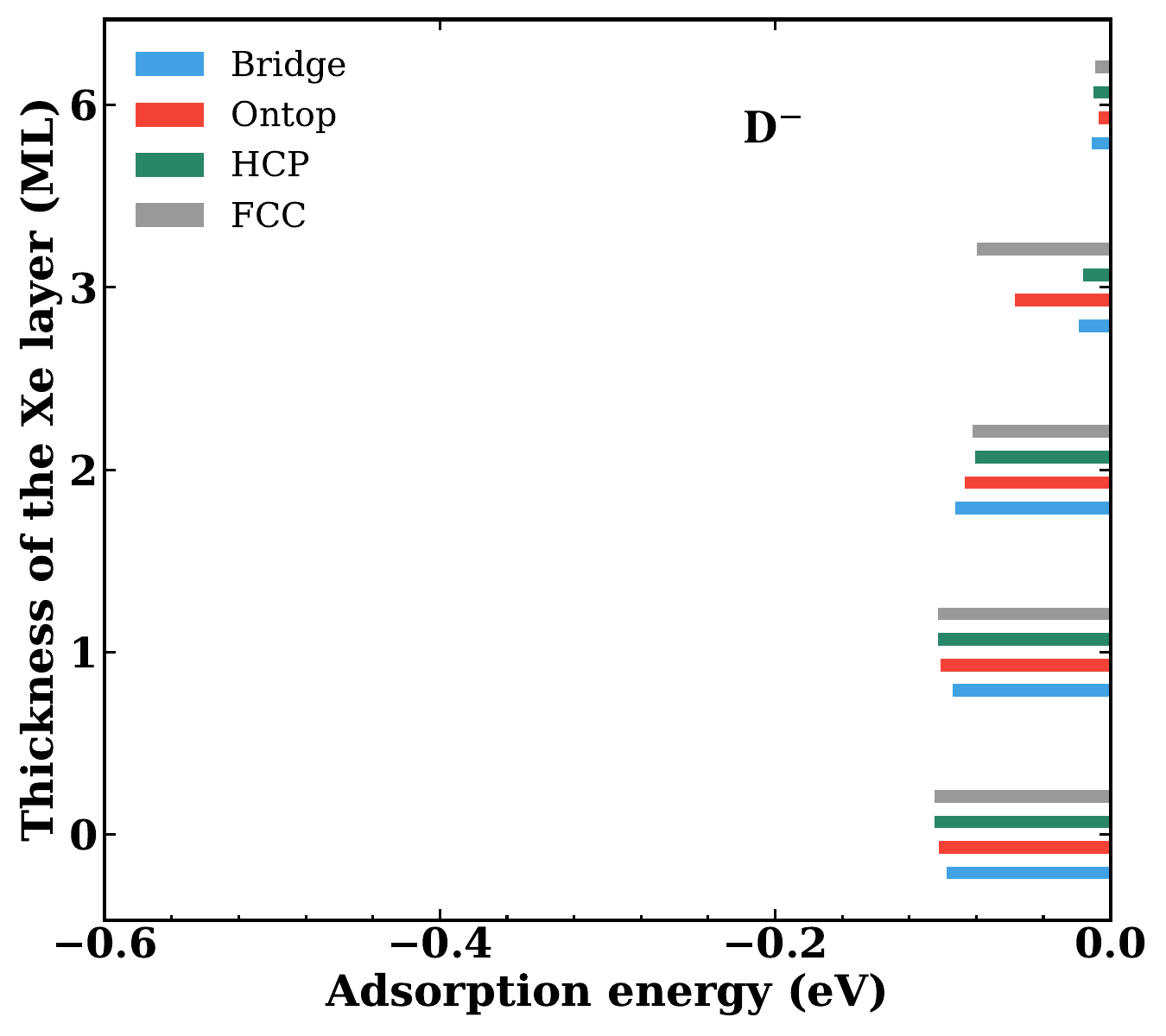}
        \label{Fig:Eads_DAn}
    \end{subfigure}
    \caption{Adsorption energy of neutral (CD$_4$ and D) and anionic (CD$_4^-$ and D$^-$) moieties as a function of Xe layer thickness. Each color corresponds to an adsorption site defined in Fig. \ref{fig:StructVASP}.}
    \label{Fig:Eads_AllFrag}
\end{figure*}

To estimate the extent of the attraction force to the substrate, we have carried out calculations of the adsorption energy for the molecular species CD$_4$, CD$_{4}^{-}$, D$^-$ and D which are present on the metal surface when D$^-$ desorption occurs. Here, we assume that D could arise from D$^-$ spontaneously loosing an electron, preferably to the metal surface. We further assume that CD$_4^-$ is a stable anion, when we calculate the adsorption energies. These suppositions avoid taking the species’ lifetime into consideration. The results are given in Fig.\ref{Fig:Eads_AllFrag}. Calculations were also performed for different adsorption sites of fragment X, including the Bridge, On-top, HCP and FCC configurations according to Fig.\ref{fig:StructVASP}b. On the bare surface of Pt, the equilibrium distance with respect to the surface of each molecular species of interest were determined prior to adsorption energy calculations. Moreover, calculations were conducted to determine the energies for different thicknesses of Xe layers, specifically for thicknesses of 0-3 and 6 ML. For D$^-$ and CD$_4$, we obtained adsorption energies characteristic of physisorbed species. These species have weak interactions with the Pt surface, i.e., E$_{ads}$ = -0.1 eV for D$^-$ and -0.4 eV for CD$_4$ at the platinum surface. Obviously, at greater thicknesses of xenon layers, the fragments are less influenced by the platinum substrate. For example, E$_{ads}$ is -0.01 eV for both fragments at 6 ML of Xe. On the other hand, D exhibits strong adsorption on the Pt(111) surface. At 0 ML of Xe, D adsorption energy is -3.9 eV, indicating a chemisorbed state. Of course, as Xe thickness increases, D adsorption energy decreases, but still reaching -2.0 eV (chemisorbed) at 3 ML and -0.5 eV (physisorbed) at 6 ML of Xe on Pt(111). Similarly, CD$_{4}^{-}$ is chemisorbed (E$_{ads}$ = -3.2 eV) on the metal surface in the absence of a xenon layer, but becomes physisorbed (E$_{ads}$ = -0.3 eV) at 6 ML of Xe on Pt. \\

\begin{figure}
\centering
\includegraphics[scale=0.3]{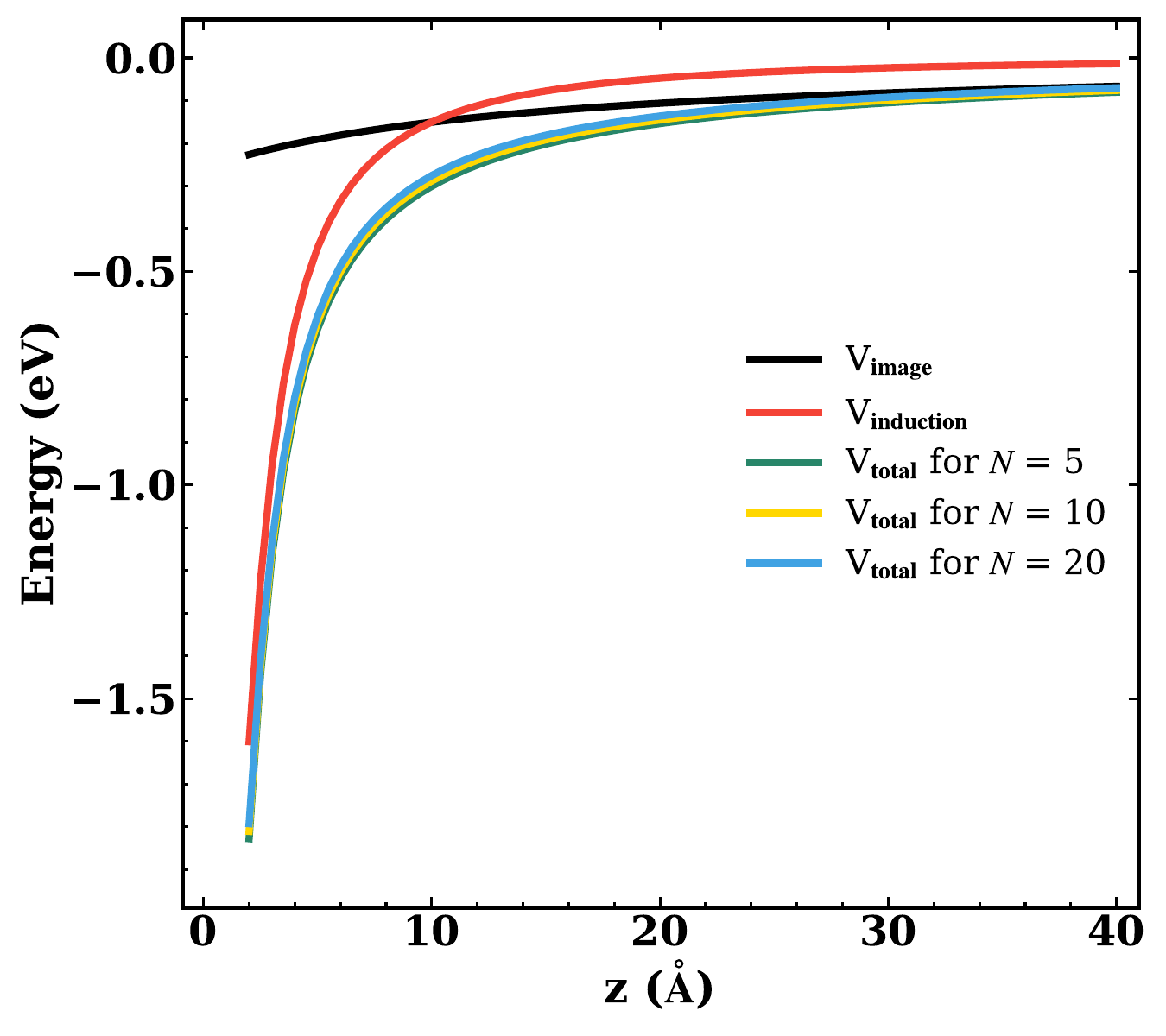}
\caption{Image potential between an anion of elementary charge and the Pt surface (black line) and induction energy for a 5 ML Xe slab (red). The total energy is also reported for different Xe layers, namely $N=5$ (green), $N=10$ (yelow) and $N=20$ (blue). The $z$ correspond to the distance between the anion and the top Xe layer.}
\label{fig:induction}
\end{figure}

In summary, the energy calculations provide several insights: 
i) CD$_4$ molecules are always physisorbed on the Pt surface and on every Xe spacer, indicating that the electronic ground and excited states of the gas phase CD$_4$ molecule are not significantly altered by the presence of the surface. ii) D$^-$ is weakly physisorbed on Pt and the Xe spacer, thus allowing the anion to desorb easily from the Pt or Xe/Pt interfaces. This contributes to the high ESD yields of D$^-$ observed experimentally. iii) Beyond 6 ML of Xe, the effect of the metallic surface on CD$_4$ and its fragments diminishes. A deformation of the resonance peak and its elongation towards higher energies is clearly visible in Figure \ref{Fig:Yields_2MLCD4} for 0-3 ML. Above 3 ML, the resonances appear free of deformation, suggesting alteration of the TA states below 3 ML. The fragments with adsorption energy values below -1 eV are mostly physisorbed. This trend is observed regardless of the fragment position and nature on the surface. The range of attraction to the substrate is thus limited to a few xenon monolayers. At that stage, this result should be taken with caution since DFT calculations using the GGA density functionals are well-known to give inaccurate description of the long range van-der-Waals interaction. More importantly, they do not reproduce the physically expected 1/r convergence of the image-charge potential to zero in the long range. Indeed, DFT is known to display a strong electron delocalization error. This latter point could not be disregarded, especially at large ion-surface distance. The instantaneous polarization effects due to induced charges in the free electron gas are in that sense much better described using a semi-empirical approach by Geada et al. \citep{Geada2018}. The image potential can be written as:
\begin{equation}
 	V_{image}  = \frac{1 } {4 \pi \epsilon_0 } \  \frac{ -  Q^2}{4(d-z_{im})} 
 	\label{Eq:Vimage}
 \end{equation}

where $\epsilon _0$ is the dielectric constant in vacuum, $Q$ is the charge of the ion, $d$ is the vertical distance of the ion with respect to the metal surface atomic plane, $z_{im}$ is the vertical coordinate of the jellium edge above the metal surface atomic plane. It takes into account the electronic cloud thickness at the metal surface and is chosen to be equal to 1 \AA \ according to Geada et al. \cite{Geada2018}. \\
We have also  calculated the induction energy occurring between the charge $Q$ and the Xe spacer. The spacer is composed of $N$ layers of Xe atoms of electronic polarizability $\alpha$ =4 \AA$^3$ arranged in a hexagonal lattice. The interlayer distance of Xe is chosen to be $h$=3.58 \AA. 
The induction energy can be written as:
\begin{equation}
    V_{induction }=-\frac{1}{4 \pi \epsilon _0} \frac{Q^2 \alpha}{2} \sum _{i=-\infty} ^{+\infty} \sum _{n=1} ^{N} \frac{1}{\left ( (z+nh)^2+x_i^2+y_i^2 \right)} 
\label{Eq:induction}
\end{equation}
 
where $x_i$ and $y_i$ refer respectively to the position of the $i^{th}$ Xe atom in the $n^{th}$ layer and $z$ the distance of this atom with respect to ontop Xe layer. It should be noted that $d=z+Nh$.    
\\
Figure \ref{fig:induction} shows the image potential due to the metal and the induction contribution due to 5 layers of Xe which corresponds to a 14.3 \AA \ slab thickness. The contribution of the image potential is less than 0.25 eV when $z$ is larger than 5 \AA, i.e. for a distance $d$ between the anion and the metal larger than 19.3 \AA. The induction energy is the dominant contribution at short $z$ distances, less than 10 \AA \ for $N=5$. This energy ranges between -1.5 to -0.5 eV for $4 < z < 5$ \AA. At larger distances, it decreases more rapidly to zero than the image potential. The total energy, sum of the image potential and induction energy, is also very instructive. It clearly shows that, for slabs larger than 5 ML of Xe, no major change is observed. The range of attraction due to the substrate and the Xe spacer is thus limited to a few xenon monolayers. This is in line with the previous mentioned results obtained by DFT.      \\

In other words, adsorption phenomena due to the metal surface or the spacer cannot explain the significant increase in anion yields observed for Xe thicknesses larger than 6 ML or the saturation occurring at more than 20 ML. Other factors must be considered. At this stage, a closer examination of the interactions between incident low-energy electrons (4-16 eV) and this dielectric layer should be conducted.

\begin{figure}[h!]
\centering
\begin{subfigure}
        \centering
        \includegraphics[scale=0.35]{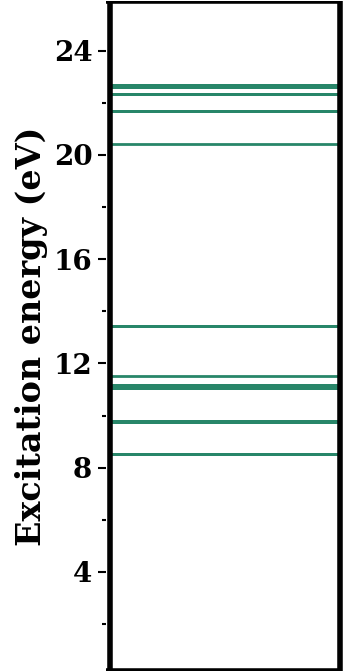} 
        \label{Fig:ES_Xe_N}
    \end{subfigure}
    \hfill%
    \begin{subfigure}
      \centering
       \includegraphics[scale=0.35]{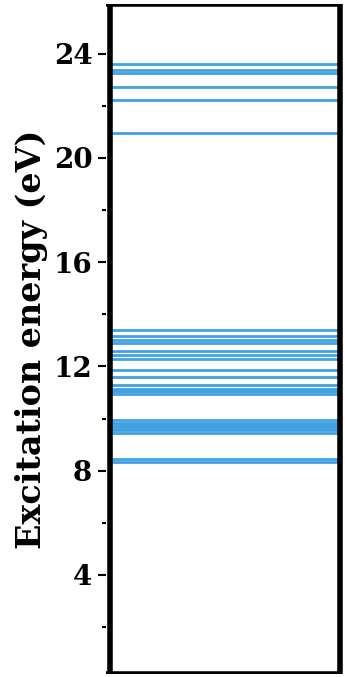} 
        \label{Fig:ES_Xe-LS}
    \end{subfigure}
    \caption{Calculated excitation energies of the neutral xenon atom (left panel) and experimental data reported in  Ref.\cite{saloman2004energy} (right panel).}
    \label{Fig:ES_Xe}
\end{figure}

The calculated excited states of the Xe are shown in Figure \ref{Fig:ES_Xe}. The first excitation energies for Xe typically appear around 8-9 eV. These energy values are consistent with the ones reported in previous studies \cite{saloman2004energy}, where the first excitation energy of Xe is observed around 8.3 eV. They belong to the energy range of the incident electrons (4-16 eV) used in our experiments and surprisingly at the same energies where CD$_4$ DEA appears to be efficient. In that case, the resonance energy closely matches these values. The collision between incident electrons and Xe atoms can thus involve either elastic or inelastic scattering, in solid rare gases. These results show that, in the energy range of 8-10 eV, if xenon atoms interact with incident electrons, the electrons can potentially be temporarily captured by Xe atoms and coupled to CD$_4$ resonance states. 
This has already been observed experimentally. Indeed, it has been shown that anionic excitons can be created by electron attachment to an atom in rare gas solids \cite{Rowntree1993,Sanche1972}. The resonance  state has a velocity vector and hence can reach adsorbed molecules, where it can couple to their dissociative Rydberg states, but not with the dissociative valence states. A core-excited resonance formed by the impact of LEE in solid xenon layers have been measured at 7.7 $\pm$ 0.2 eV \cite{Rowntree1993}. According to the resonance width the lifetime of the 7.7 eV Xe$^{-*}$ anion is longer than 10 femtoseconds \cite{Sanche1972}, which is sufficient time for the anion state to diffuse to the surface \cite{Rowntree1993}. Moreover, this core-excited TA has by far the largest capture cross section of all possible anionic states resulting from electron attachment to Xe \cite{Sanche1972}. Thus, formation of this TA can redistribute electrons in direction of the metal to be redistributed via autoionisation in all directions in the Xe film.
This could result in a higher number of electrons reaching the CD$_4$ target molecules, which in turn increases the likelihood of the CD$_4$ molecules capturing an electron. Alternatively, the electron-exciton complex formed at 7.7 eV could reach  surface molecules, where it could exchange energy and charge with CD$_4$ to create a dissociative CD$_4^-$ state and produce D$^-$ \cite{Rowntree1993}. While these possibilities are invoked taking as an example the lowest energy resonance of Xe, we expect these phenomena to be effective with the multitude of core-excited TAs covering the 7.7 to 11.5 eV range in Xe \cite{Sanche1972}, i.e., within the range of the broad DEA peaks seen in Figure \ref{Fig:Yields_2MLCD4}. \\

Anionic desorption after dissociation of the TA is closely linked to the kinetic energy distributed among the fragments produced during the process. As the thickness of the xenon layer increases, the adsorption energy on the surface diminishes, leading to a decrease in the attractive force exerted on the stable anion formed within the film. Therefore, the fragments become less adsorbed and require less kinetic energy to desorb from the film. This finding may explain the experimental increase in anion desorption with increasing xenon layer thickness. However, our results indicate that the potential well reaches a minimum starting from a thickness of 6 ML (where all the fragments are physisorbed), while experimental results indicate a desorption saturation around 20 ML (Fig.\ref{fig:DYield_Max}).
Two hypotheses can explain the saturation phenomena observed exclusively from 20 ML of xenon: on one hand, it is possible that the number of electrons reflected towards the CD$_4$ target molecules and the electron-exciton path reach their maxima, and on the other hand, it is possible to reach a maximum level of DEA by isolating the metal surface with Xe layers while saturating the induced polirization. 

\section{Conclusion}

Based on the information provided, it seems that the Xe spacer is a crucial component in enhancing the efficiency of the DEA process in CD$_4$. The Xe spacer serves as a barrier between the platinum surface and CD$_4$, reducing the effect of substrate adsorption, which can interfere with the DEA process, e.g., by reducing the CD$_4^-$ anion lifetime and the velocity of D$^-$. The experimental results show that increasing the thickness of the Xe spacer results in a higher yield of fragments produced by DEA to CD$_4$. This indicates that the Xe spacer plays a critical role in promoting DEA. Moreover, calculations show that beyond 6 ML of Xe on the platinum surface, the influence of the substrate vanishes, implying that at higher Xe thicknesses, all anionic and neutral fragments produced are physisorbed and interact weakly with the substrate. This finding further supports the importance of the Xe spacer in reducing the effect of substrate adsorption. Additionally, calculations of Xe-excited states provide evidence for the existence of well-localized electronic states in the incident electron energy range of the present measurements. This finding suggests that the Xe spacer can trap low-energy incident electrons by forming electron-exciton TA states, thus promoting attachment to CD$_4$ and resulting in an increase in the yield of DEA fragments. Overall, the Xe spacer plays a crucial role in promoting the efficiency of the DEA process in CD$_4$ by reducing the effect of substrate adsorption and serving as an electron reservoir that promotes temporary attachment to CD$_4$.

\begin{acknowledgments}
We acknowledge the facilities that offered the essential support for conducting the computations in this study. This support was provided by the Mésocentre de Franche-Comté and the TGCC (Très Grand Centre de Calcul du CEA). We were able to access the TGCC resources thanks to the allocation 2021-A0110813008 awarded by GENCI. Financial support for the experiments was provided by the Natural Science and Engineering Research Council of Canada (RGPIN-2018-14882).
\end{acknowledgments}

\section*{Conflict of interest}
The authors have no conflicts to declare.

\nocite{*}
\bibliography{aipsamp}

\end{document}